\documentclass[twocolumn, secnumarabic,amssymb, nobibnotes, aps, prd, superscriptaddress]{revtex4-1} 
\usepackage{amsmath}
\usepackage[breaklinks=true]{hyperref}
\usepackage{cleveref}

\usepackage{graphicx}
\usepackage[font=normal]{subfig}

\usepackage{booktabs}
\usepackage{tabularx}

\usepackage{enumitem}
\setlist[itemize]{align=parleft,left=0pt..1em}

\usepackage[dvipsnames]{xcolor}

\begin{document}

\title{Variational Quantum Optimization with Multi-Basis Encodings}

\author{Taylor L. Patti}
\email[]{taylorpatti@g.harvard.edu}
\affiliation{Department of Physics, Harvard University, Cambridge, Massachusetts 02138, USA}
\affiliation{NVIDIA, Santa Clara, California 95051, USA}

\author{Jean Kossaifi}
\affiliation{NVIDIA, Santa Clara, California 95051, USA}

\author{Anima Anandkumar}
\affiliation{Department of Computing + Mathematical Sciences (CMS), California Institute of Technology (Caltech), Pasadena, CA 91125 USA}
\affiliation{NVIDIA, Santa Clara, California 95051, USA}

\author{Susanne F. Yelin}
\affiliation{Department of Physics, Harvard University, Cambridge, Massachusetts 02138, USA}

\begin{abstract}
Despite extensive research efforts, few quantum algorithms for classical optimization demonstrate realizable quantum advantage. The utility of many quantum algorithms is limited by high requisite circuit depth and nonconvex optimization landscapes. We tackle these challenges by introducing a new variational quantum algorithm that benefits from two innovations: multi-basis graph encodings and nonlinear activation functions. Our technique results in increased optimization performance, a factor of two increase in effective quantum resources, and a quadratic reduction in measurement complexity. While the classical simulation of many qubits with traditional quantum formalism is impossible due to its exponential scaling, we mitigate this limitation with exact circuit representations using factorized tensor rings. In particular, the shallow circuits permitted by our technique, combined with efficient factorized tensor-based simulation, enable us to successfully optimize the MaxCut of the nonlocally connected $512$-vertex DIMACS library graphs on a single GPU. By improving the performance of quantum optimization algorithms while requiring fewer quantum resources and utilizing shallower, more error-resistant circuits, we offer tangible progress for variational quantum optimization.
\end{abstract}

\maketitle

\section{Introduction}

\begin{figure*}
\subfloat[\textbf{Multi-Basis Encoding (MBE) of a graph}. An $n$-vertex graph (blue) is represented as an Ising model. We reassign $n/2$ vertices from $\sigma^z$ (blue) to $\sigma^x$ (red) operators, allowing us to map the graph to just $n/2$ qubits (here a nearest-neighbors connected, blue/red tensor ring). The MaxCut is obtained by optimizing this state via single-qubit measurements. Although only locally connected, tensor rings effectively solve the MaxCut graphs with highly nonlocal connections. \label{fig.4a}]{%
  \includegraphics[width=0.95\columnwidth]{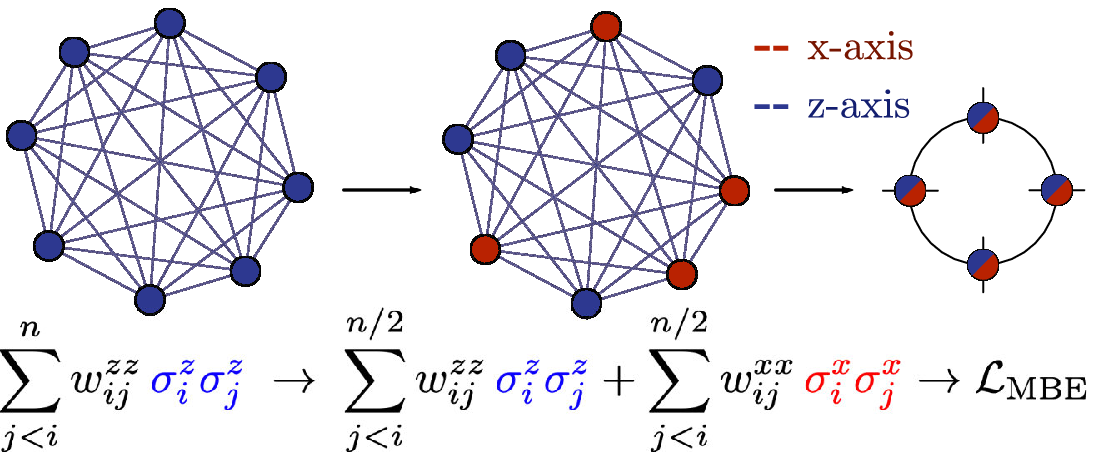}%
}
\hspace{1cm}
\subfloat[\textbf{Multi-Basis Encoding (MBE) with two distinct $n$-qubit graphs.} Each graph is mapped to the classical Ising model, with $G_0$ (blue) encoded along the $z$-basis (as is traditional) and $G_1$ (red) utilizing the $x$-basis, resulting in an $n$-qubit quantum state (blue/red). This encoding is similar to MBE with a single graph, except that the $x$ and $z$-bases \textit{independently} encode two separate graphs and thus no cross-terms between the $z$ and $x$-bases are required. \label{fig.2a}]{%
  \includegraphics[width=0.95\columnwidth]{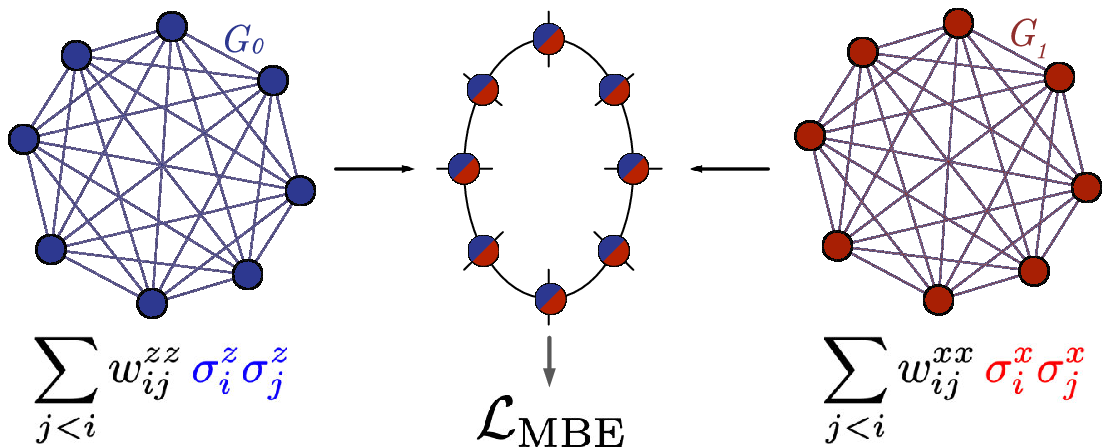}
}
\caption{}
\label{}
\end{figure*}

NP-hard optimization problems, such as Traveling Salesman and MaxCut, are central to a wide array of fields, such as operational research, engineering, and network design~\cite{Li2020}. Despite the classical nature of these problems, there is immense interest in identifying variational quantum algorithms (VQAs) which can solve them faster or more precisely than any classical method, a concept known as quantum advantage~\cite{Lucas2014,wecker2015,McClean_2016,cerezo2020variational}.

One common approach is the variational quantum eigensolver (VQE), where energy minimization yields the ground state of a problem-encoded Hamiltonian through gradient descent update of the quantum circuit parameters~\cite{Peruzzo2014,Kandala2017,lee2021favorable}. The quantum approximate optimization algorithm (QAOA) is a related protocol in which unitary evolutions using both an initial and a problem encoded Hamiltonian are alternated in order to find a solution encoded ground state~\cite{Farhi2014,Harrigan2021,Guerreschi2019,Pagano,Zhou2020Leo}. Novel VQA encoding strategies have also been considered in~\cite{kim2017robust,Wang2020Rieffel,Fuchs2021}. While the approximation ratio of VQE can surpass those of polynomial complexity classical algorithms (e.g., Goemans-Williamson~\cite{GW,Hastad2001,Khot2005})~\cite{lee2021favorable}, this guarantee requires between polynomially and exponentially many gates in the number of qubits $n$. Such circuit depths limit the algorithms' potential to demonstrate quantum advantage, rendering them not only computationally inefficient, but also highly susceptible to quantum noise \cite{Preskill2018quantumcomputingin,Guerreschi2019,Harrigan2021} and barren plateaus \cite{McClean2018,Patti2020,Marrero2020, Wiersema2020,holmes2021connecting,Cerezo2020}. Moreover, local VQAs, where quantum state update is limited to only explicitly connected degrees of freedom, have demonstrably poorer performance than classical methods on particularly challenging and large graph instances \cite{Hastings2019, Bravyi2020}.

The difficulty of classically simulating large-scale quantum circuits is a central challenge to algorithm development. This is because the traditional mathematical formalism of quantum mechanics automatically represents the full Hilbert space and thus scales exponentially in the number of qubits $n$, with matrix operators of size $2^{2n}$ operating on state vectors of size $2^n$. When a quantum system does not occupy the full Hilbert space, these intractable dimensions for quantum network simulation can be remediated by employing a factorized tensor formalism~\cite{fishman2020itensor}. While many varieties of decomposed tensors exist, tensor rings have proven particularly popular in the quantum sciences due to their modularity and rank structure, which have close parallels to quantum entanglement. In the tensor ring formalism, both quantum states and quantum operators are represented in factorized form by matrix product states (MPS) and matrix product operators (MPOs), respectively~\cite{Orus,Bridgeman,Huggins2019}. However, tensor formalism is often unsuitable for high-depth and connectivity regimes, which are most commonly used in quantum optimization, since tensor rings quickly become prohibitively large (high-rank/bond-dimension) when simulating deep or complicated circuits \cite{Zhou2020}. Moreover, they are limited to only nearest-neighbor interactions.

Due in part to these limitations, no simulation of more than $\sim 100$ qubits has demonstrated quantum optimization rivaling that of classical methods for nonlocally connected graph instances, even in \cite{Fried} where exact representations of general tensor architectures with optimal contraction schemes are used. Other large-scale implementations have focused on more restrictive problems. For instance, QAOA MaxCut optimization with up to $210$ qubits has been achieved for $3$-regular graphs with nonlocal edges~\cite{lykov2020}. QAOA MaxCut optimization has also been implemented with several thousand qubits when exploring only local edges of nonlocally connected graphs, a method which did not yield high average performance \cite{huang2019}. Moreover, large-scale optimization on NP-hard problems (e.g., MaxCut) have not been explored using VQE.

\textbf{Quantum Computing Contribution -} This manuscript introduces a novel method for quantum algorithms that not only outperforms traditional VQAs, it also requires fewer quantum resources and lower computational complexity. In particular:

\begin{itemize}

\item We propose Multi-Basis Encodings (MBEs), a new quantum optimization algorithm that introduces additional constraints (regularization) that are \textit{beneficial} to the algorithm's performance, reducing its susceptibility to local minima in the training landscape.

\item By doubling the amount of optimization features encoded into a single qubit, MBEs halve the number of qubits required for a given optimization task, a valuable asset for a developing field which has invested millions of dollars and spent multiple decades to achieve $\sim 50$-qubit registers and where additional coherence limitations emerge at scale \cite{deLeon2021}. Moreover, by utilizing single-qubit measurements, these algorithms yield up to a quadratic reduction in runtime.

\item By combining our MBEs with non-linear activation functions and an exact factorized tensor network approach, we solve MaxCut graph optimization problems with nonlocal edges using shallow quantum circuits. Furthermore, sampling $\sim 5$ initializations of our MBE experiments on shallow circuits (depth $L=7$ for $100$-vertex graphs, such that $L$ approximately logarithmic in the number of vertices) leads to optimal cut convergence with near unit probability. This shallow-circuit, multi-shot procedure is both more coherent and time-efficient than deterministic convergence with deep circuits, which require up to an exponential number of parameters.

\end{itemize}

\textbf{Large-Scale Simulation Contribution - } This work utilizes tensor networks, developing new software in order to simulate practical quantum algorithms at unprecedented scale. Specifically:

\begin{itemize}

\item The strong performance of our MBE with relatively shallow circuits enables us to work with tensor networks with lower rank (bond dimension). As the rank of a tensor structure determines the time and memory complexity of its contraction, we can simulate high-accuracy implementations of MBE at large scales.

\item We develop TensorLy-Quantum~\cite{Patti2021TLQ, TLQ_URL}, a new software package for simulating efficient quantum circuits with decomposed tensors on CPU and GPU. TensorLy-Quantum is based on the TensorLy software family~\cite{Kossaifi2019}.

\item Using TensorLy-Quantum on a single NVIDIA A100 GPU, we simulate solving a $512$-vertex MaxCut problem using MBE, which demonstrates superior performance than comparable classical algorithms. This sets a new record for the large-scale simulation of a successful quantum optimization algorithm.
\end{itemize}

\noindent By introducing a new variety of algorithms that improve optimization performance, require fewer quantum resources, and operate on shallower, more error-resistant circuits, we offer tools to increase the utility of variational quantum algorithms.

\begin{figure*}
\includegraphics[width=1\textwidth]{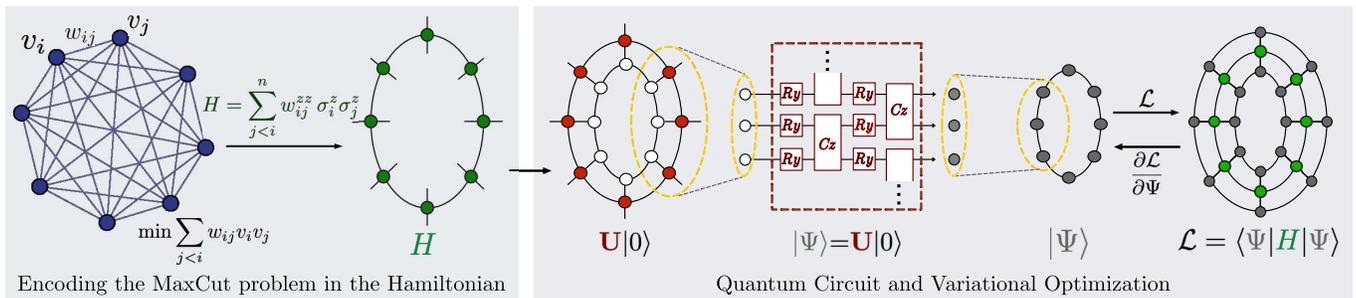}
\caption{\textbf{Overview of traditional MaxCut encoding and VQE using tensor ring factorizations}, which are tensor train networks with periodic boundary conditions. (Left) A graph $G$ with $n$ vertices $v_i$, $v_j$ and weights $w_{ij}$ is mapped into an $n$-qubit Hamiltonian $H$ in MPO form. The MPS ground state $|\psi_g\rangle$ of $H$ encodes the solution to MaxCut$(G)$. (Right) To find MaxCut$(G)$ variationally, the null input state $|\mathbf{0}\rangle$ (an MPS) is evolved under a parameterized quantum circuit $U$ (an MPO), producing an output state $|\psi \rangle$. $U$ encodes a circuit of depth $L$ (here $L=4$, red box) in this manuscript's layer (block) pattern: one layer (block) of single-qubit $y$-axis rotations $R_y$ followed by a layer of control-Z gates which alternate between even and odd qubits. The energy expectation value $\mathcal{L} = E$ is minimized via gradient descent. The global minimum of $\mathcal{L}$ corresponds to $|\psi\rangle = |\psi_g\rangle$.}
\label{fig.1}
\end{figure*}

\subsection{MaxCut Optimization Problems}

The Maximum Cut problem, most commonly referred to as \emph{MaxCut}, is a partitioning problem on unidirected graphs $G = (V,A)$, where $V$ is a set of vertices (blue orbs in Fig.\ \ref{fig.1}, left) connected by edges $A$ (black lines connecting orbs)~\cite{Commander2009}. The objective is to optimally assign all vertices $v_i$, $v_j \in \{-1,1\}$, so as to maximize the edge weights $w_{ij} \in A$, where any such assignment is referred to as a ``cut". In this work, we will consider a generalized form of the problem known as \textit{weighted} MaxCut, in which $w_{ij}$ take arbitrary real values.

Two formulations of MaxCut exist: the NP-complete decision problem and the NP-hard optimization problem~\cite{Karp1972}. The former seeks to determine if a cut of size $c$ or greater exists for a given graph $G$, whereas the latter attemps to identify the largest cut of $G$ possible. We here focus on the more general optimization problem formulation, the ground truth of which we denote MaxCut$(G)$. It is common practice to express the objective function in its binary quadratic form~\cite{Commander2009}:

\begin{equation}
\textrm{maximize} \hspace{0.4cm} \frac{1}{2} \sum_{j<i} w_{ij} \left(1-v_i v_j \right).
\label{eq.MaxCut}
\end{equation}

\subsection{VQE Framework and Tensor Network Formalism}

To find the MaxCut of a given graph on a quantum computer, it is convenient to minimize the equivalent summation, $\sum_{j<i} w_{ij} v_i v_j$. For a graph with $n$ vertices $v_i$, this reduces the problem to finding the $n$-qubit wavefunction $| \psi \rangle$ that minimizes the energy expectation value $E = \langle \psi | H | \psi \rangle$ of the classical Ising Model Hamiltonian:
\begin{equation}
H = \sum_{j<i}^{n} w_{ij}^{zz} \sigma_i^z \sigma_j^z.
\label{eq.HIsing}
\end{equation}

\noindent $H$ is obtained by substituting vertices $v_i$ for the Pauli-Z spin operators $\sigma_i^z$, as depicted in Fig.\ \ref{fig.1}, and $w_{ij}^{zz}=w_{ij}$ is a relabeling to specify the $zz$-spin interactions. As $H$ contains only terms in the $z$-basis, its eigenvectors are classical (zero-entanglement product states), such that $|\psi_i \rangle = \bigotimes_{s}\,|s\rangle$, where $|s\rangle \in \{|0\rangle, |1\rangle \}$. We here denote the lowest eigenvalue or ``ground state" solution as $|\psi_g\rangle$, the qubits of which form a bijection with the optimal $v_i$ of MaxCut$(G)$. As Eq.~\ref{eq.HIsing} has $\mathbb{Z}_2$ symmetry, $|\psi_g\rangle$ is degenerate with the state $X^{\otimes n} |\psi_g\rangle$.

Fig.\ \ref{fig.1} (right) depicts the VQE framework \cite{Peruzzo2014, Kandala2017, lee2021favorable}. Eq.\ \ref{eq.MaxCut} is optimized by defining the loss function $\mathcal{L} = E$ and varying the parameters $\hat{\theta}$ of a quantum circuit with unitary $U(\hat{\theta})$, which acts on the input quantum state (Fig.\ \ref{fig.1}, right). Without loss of generality, we define the input state as the $n$-qubit zero state $|\mathbf{0}\rangle = \bigotimes_{n}\,|0\rangle$, such that

\begin{equation}
|\psi\rangle = U(\hat{\theta}) |\mathbf{0}\rangle.
\end{equation}

\noindent We decompose this unitary matrix $U$ as $\Lambda$ subunitaries $U(\hat{\theta}) = \prod_k^{\Lambda} U_k(\hat{\theta}_k)$, where $\hat{\theta}_k$ is the corresponding subset of $\hat{\theta}$ and $U_k(\hat{\theta}_k) = \prod_{j=1}^n \exp(-i\hat{\theta}_j W_j) M_k$ for generic Hermitian operators $W_j$ and unitary matrices $M_k$. Thus, the gradient $g_l(\hat{O}) = \frac{\partial \langle \hat{O} \rangle}{\partial \theta_l}$ of operator $\hat{O}$ with respect to any parameter $\theta_l \in \hat{\theta}$ is

\begin{equation}
g_l(\hat{O}) = i\langle \mathbf{0} | U_R ^\dagger \left[W_l, U_L^\dagger \hat{O} U_L \right] U_R | \mathbf{0} \rangle,
\label{eq:gradient}
\end{equation}

\noindent where $U_L$ and $U_R$ are the compositions of unitaries $U_k$ with $k \geq l$ and $k < l$, respectively. Rather than using circuits with extensive connectivity, we instead focus on 1D tensor ring circuits of $n$ qubits. In particular, tensor rings have periodic boundary conditions such that qubit $n-1$ is connected to qubit $0$. Such nearest-neighbor connectivity makes the circuit amenable to both near-term quantum hardware \cite{Harrigan2021, Pagano} and simulation via decomposed tensors. We accomplish this simulation with TensorLy-Quantum \cite{Patti2021TLQ, TLQ_URL}. A nascent and expanding software package, TensorLy-Quantum strives to leverage the structure of decomposed tensors in order to simulate quantum machine learning in the most efficient, non-approximate manner possible. While tensor ring-based tensor networks are typically used for approximate inference and obtained by applying tensor decomposition to dense state vectors and operators, we build a low-rank but exact factorized representation of the simulated quantum circuits. When judiciously constructed, tensor simulations yield a low-rank quantum formalism that permits enormous compression of state and operator spaces. Although in the quantum sciences tensor methods are most frequently associated with state approximations and truncations, like the density matrix renormalization group \cite{White1992}, we here advocate for their use in exact quantum simulation. Similarly, due to their nearest-neighbor connectivity, tensor ring factorizations in quantum computing have traditionally been employed for locally connected optimization problems, such as $3$-regular MaxCut \cite{Wurtz2021}, however we here emphasize their utility for general purpose optimization tasks.

To analyze VQE with tensor formalism, the Hamiltonian of Eq.\ \ref{eq.HIsing} is represented as an MPO $H^{\{\beta,\gamma\}}$, with physical indices $\beta$ and $\gamma$. The energy $\mathcal{L}=E$ is then calculated with a single large contraction (Fig.\ \ref{fig.1}, right)

\begin{equation}
E = \sum_{\{\beta, \gamma, \delta, \epsilon\}} \Psi^{\{\beta\}} U^{\{\beta, \gamma\}} H^{\{\gamma,\delta\}} U^{\{\delta,\epsilon\}} \Psi^{\{\epsilon\}},
\end{equation}

\noindent where

$$\Psi^{\{\beta\}} = \Psi^{\beta_0,...,\beta_{m-1}} = \sum_{\{\alpha\}} \psi^{\beta_0}_{\alpha_0 \alpha_1},...,\psi^{\beta_{m-1}}_{\alpha_{m-1} \alpha_0}$$

\noindent is an $n$-qubit MPS of $m$ cores and

$$U^{\{\beta, \gamma\}} = U^{\beta_0,\gamma_0,...,\beta_{m-1},\gamma_{m-1}} = \sum_{\{\alpha\}} u^{\beta_0,\gamma_0}_{\alpha_0,\alpha_1},...,u^{\beta_{m-1},\gamma_{m-1}}_{\alpha_{m-1},\alpha_0}$$

\noindent is the corresponding MPO unitary.

As we work in the absence of quantum noise, states $|\psi \rangle$ display time-reversal symmetry and can be fully expressed with real numbers \cite{Shen2020}. We thus restrict our rotations to those of the Pauli-Y generator $\sigma^y$ and implement a simple, repeating subunitary pattern of two layers, also known as blocks. The pattern is illustrated in Fig.\ \ref{fig.1} (right): a row of parameterized single-qubit rotations $R_y(\theta)$ ($W=\sigma^y$) is followed by a row of control-z (CZ) gates, with the latter alternating control between even and odd qubits. As each single qubit rotation is a $2\times2$ dense matrix and each two-qubit control-z gate is a rank-2 MPO of two, eight-element cores, the memory requirements of the uncontracted circuit representation scale only linearly in both $n$ and $L$, an exponential reduction in resources compared to circuits described in traditional quantum formalism. Likewise, a factorized representation of the input state $|\mathbf{0}\rangle$ in tensor ring form requires exponentially fewer terms, as it is represented by a rank-$\prod_{i=0}^n 1$ MPS with just $n$, two-element cores.

\section{Multi-Basis Encoding (MBE)}

\textbf{Intuition -} Our MBE protocol uses a loss function which is inspired by, but not equivalent to, the long-range, ZX Hamiltonian

\begin{equation}
H_{zx} = \sum_{j < i} w_{ij}^{zz} \sigma_i^z \sigma_j^z + \sum_{j < i} w_{ij}^{xx} \sigma_i^x \sigma_j^x + \sum_{i,j} w_{ij}^{zx} \sigma_i^z \sigma_j^x.
\label{eq.HHZ}
\end{equation}

\noindent The key difference between Eq.\ \ref{eq.HHZ} and MBE is that MBE utilizes the product of \textit{single-qubit} measurements and nonlinear activation functions to encode separate vertices into the $z$ and $x$-bases (further explained in Eqs.\ \ref{eq.Ldense} and \ref{eq.Cdense}). The utilization of two, rather than a single, quantum basis has proven useful in other quantum machine learning algorithms \cite{Gao2021}.

\textbf{Algorithm -} MBE for weighted graphs is depicted in Fig.\ \ref{fig.4a}. An $n$-vertex graph $G$ is expressed similarly to the Ising model Hamiltonian in Eq.\ \ref{eq.HIsing}, save that only the first $\text{ceil}(n/2)$ vertices are mapped to the $z$-axis (blue), while the second $\text{floor}(n/2)$ vertices are mapped to the $x$-axis (red), thus enabling $n$ vertices to be encoded into only $\text{ceil}(n/2)$ qubits. If $n$ is odd, then the $x$-axis of the $n$th qubit is unneeded. It is absent from the loss function and can go unmeasured. In future work, more sophisticated vertex partitionings can be explored, such as mappings that reflect graph topology. MBE halves the number of qubits required for a given optimization, providing a meaningful decrease in quantum hardware overhead.

\begin{figure}[t!]
\subfloat[Right: Average cut $\mathcal{C}$ convergence (left) for both MBE (solid lines) and traditional VQE (dashed) with $L=7$ ($n=8, 100$) and $L=13$ ($n=512$). We note the significantly increased performance for $n=8,100$ with MBE over VQE. While VQE with $n=512$ was prohibitively memory inefficient to simulate for comparison, MBE with $n=512$ outperforms VQE with $n=8$, a system $1/64$th of its size, as well as the leading single-shot classical algorithm (Table \ref{tab:table2}). Left: Average entanglement entropy for two-qubit subpartitions (maximum value per qubit is $1$) vs fraction of calculated MaxCut convergence for nonlinear loss functions. Product state formation occurs because minimizing $\mathcal{L}_\text{MBE}$ maximizes $\langle \sigma_i^z \rangle^2 + \langle \sigma_i^x \rangle^2$. Inset: $\tanh(x)$ nonlinear activation function further disincentivizes the maximization of one axis at the expense of the other. \label{fig.4b}]{%
  \includegraphics[width=1.05\columnwidth]{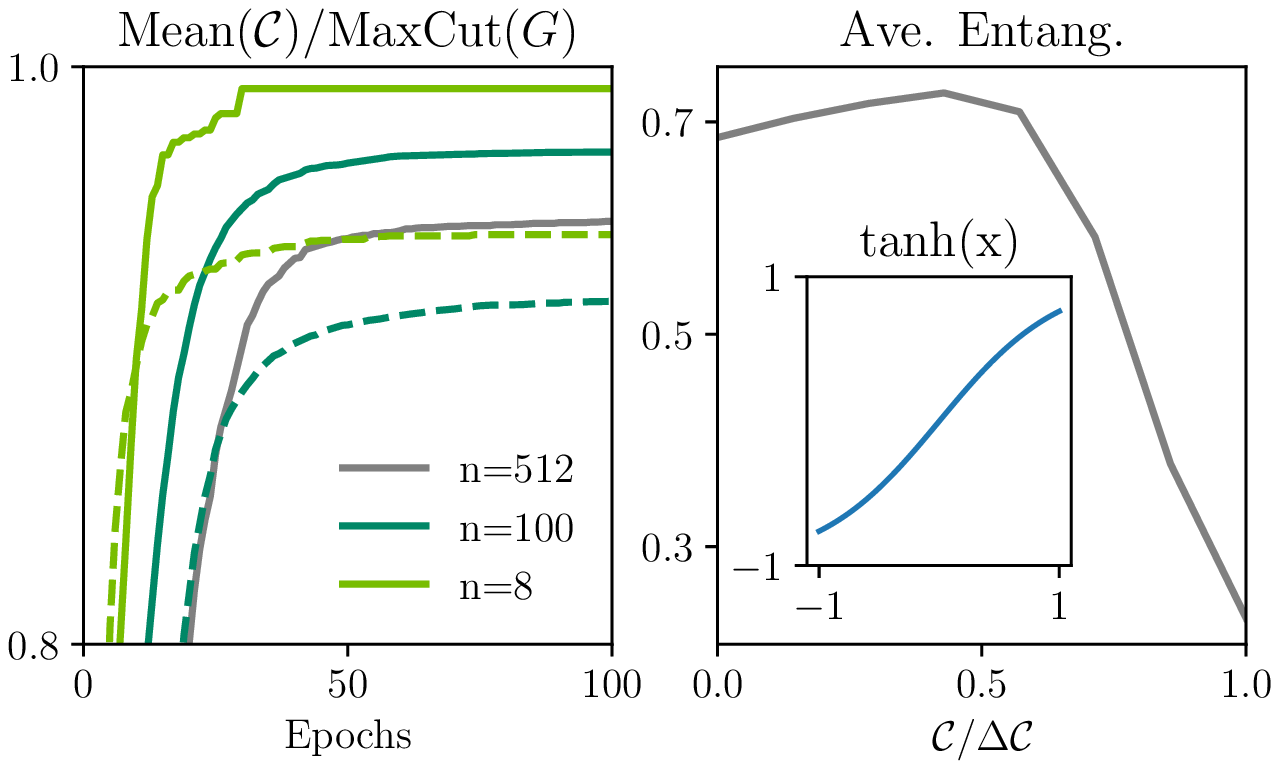}%
}\hfill
\subfloat[Average cut $\mathcal{C}$ convergence (left) and raw loss function $\mathcal{L}$ (right) with both two-graph MBE (solid lines) and traditional VQE (dashed) for $n=8$, $20$, and $100$. MBE improves \textit{calculated} MaxCut convergence $\mathcal{C}$, although its ability to satisfy by the two encoded Ising models is limited by the normalization condition of Eq.\ \ref{eq.normalization}. This is remedied by the rounding proceedure of Eq.\ \ref{eq.Cparallel}. \label{fig.2b}]{%
  \includegraphics[width=1.05\columnwidth]{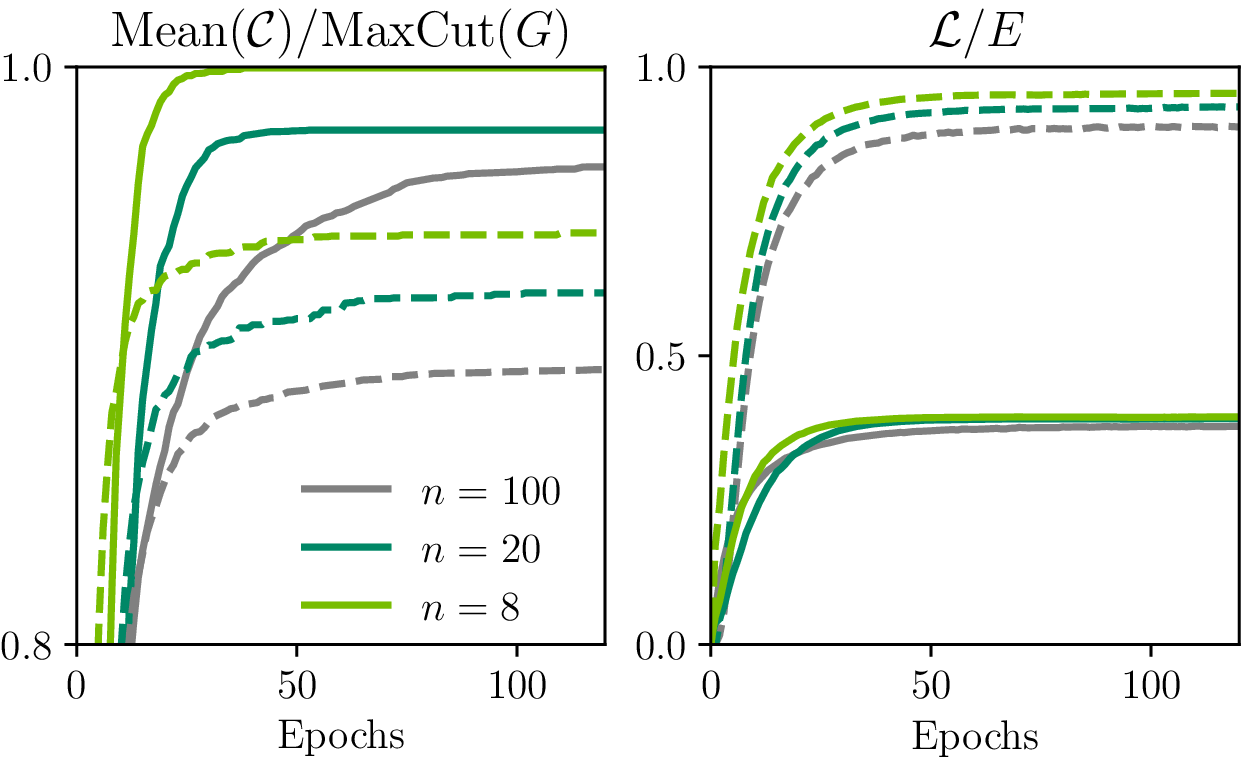}%
}
\caption{}
\label{}
\end{figure}

In order to optimize both axes as independent vertices, we must make several alterations to standard VQE. To begin, $\langle H_{zx} \rangle$ itself is an unsuitable loss function, as the quantum ground state it encodes does not correspond to classical MaxCut of $G$. We instead focus on the products of single-qubit measurements $\langle \sigma_i^x \rangle $ and $\langle \sigma_i^z \rangle $, such that $\sigma_i^x$ and $\sigma_i^z$ operators are simultaneously optimized. This yields the MBE loss function

\begin{equation}
\begin{split}
& \mathcal{L}_\text{MBE} = \sum_{j < i}^{n/2} w_{ij}^{zz} \tanh(\langle \sigma_i^z \rangle ) \tanh(\langle \sigma_j^z \rangle ) \\ & + \sum_{j < i}^{n/2} w_{ij}^{xx} \tanh(\langle \sigma_i^x \rangle ) \tanh(\langle \sigma_j^x \rangle ) \\ & + \sum_{i,j}^{n/2} w_{ij}^{zx} \tanh(\langle \sigma_i^z \rangle ) \tanh(\langle \sigma_j^x \rangle ),
\label{eq.Ldense}
\end{split}
\end{equation}

\noindent where $\tanh(x)$ is trivially implemented on the classical computer controlling gradient descent. For example, the four-vertex graph with four-qubit Ising model encoding
$$H = \omega_{12}\sigma^z_1\sigma^z_2 + \omega_{34}\sigma^z_3\sigma^z_4 + \omega_{13}\sigma^z_1\sigma^z_3,$$
would be optimized with the two-qubit MBE loss function
\begin{equation*}
\begin{split}
& \mathcal{L}_\text{MBE} = w_{12}^{zz} \tanh(\langle \sigma_1^z \rangle ) \tanh(\langle \sigma_2^z \rangle ) \\ + w_{12}^{xx} \tanh & (\langle \sigma_1^x \rangle ) \tanh(\langle \sigma_2^x \rangle ) + w_{11}^{zx} \tanh(\langle \sigma_1^z \rangle ) \tanh(\langle \sigma_1^x \rangle ).
\end{split}
\end{equation*}
We again emphasize that, as Eq.\ \ref{eq.Ldense} is comprised of distinct Pauli strings that are independently measured on separate circuit preparations, the uncertainty principle is not violated for $w_{ij}^{zx}$ with $j=i$. The projection of high-dimensional quantum data into a lower-dimensional representation has also been explored in \cite{Eddins2021, Matty2021}. The inclusion of the non-linear activation function $\tanh(x)$ disincentives the extremization of one basis at the expense of another, which could otherwise occur because the optimal values of both $\sigma_i^x$ and $\sigma_i^z$ cannot be linearly encoded by a single quantum state due to the normalization condition of the Bloch sphere of each qubit $i$

\begin{equation}
\langle \sigma_i^z \rangle^2 + \langle \sigma_i^x \rangle^2 \leq 1,
\label{eq.normalization}
\end{equation}

\noindent where equality holds for real-valued pure states. As the gradient of $\tanh(x)$ reduces near the $\pm 1$ poles (inset Fig.\ \ref{fig.4b}), full optimization of one axis at the expense of the other is discouraged and optimal cuts are deduced by a rounding procedure (detailed below), which assigns integer vertex values but does not affect parameter update or the normalization condition of Eq.\ \ref{eq.normalization}. In this manner, MBE is a dual-axis quantum analog to linear programming relaxations \cite{Aardal1996}. Furthermore, the normalization constraint of Eq.\ \ref{eq.normalization} means that $\mathcal{L}_\text{MBE}$ can only ever partially descend into local minima and is better equipped to escape their regions of attraction. The robustness of MBE against local minima can be understood through its use of global optimization \cite{Hastings2019, Bravyi2020}, including the global optimization of single-qubit states and the dependence of the $x$-encoded vertex on a generally unconnected $z$-encoded vertex. Finally, we note that we have for simplicity neglected both external fields and $y$-basis interactions in Eq.\ \ref{eq.Ldense}, however the addition of $y$-basis terms could immediately be used to both improve the algorithm's performance, as well as to simultaneously optimize three (rather than two) graph vertices.

\begin{table*}[htbp]
\caption{\label{tab:table1}
\textbf{Comparison of single-graph MBE and traditional VQE} for $n=100$ vertex graphs for circuits of depth $L=7$. MBE requires half the number of qubits and parameters as VQE, yet produces significantly better solutions (higher cut $\mathcal{C}$), both on average and with higher probability.
}
\centering
\begin{ruledtabular}
\begin{tabular*}{1\textwidth}{@{\extracolsep{\fill}} ccccccc}
\toprule
\textbf{Method}&
\textbf{Depth} &
\textbf{\#Vertices}&
\textbf{\#Qubits}&
\textbf{\#Param}&
\textbf{Mean($\mathcal{C}$)/MaxCut($G$)}&
$P(\mathcal{C} > T)$
\\
\midrule
\textbf{VQE}        & $L=7$ & 100 & 100 & 400 & 0.921 & 12.5\%\\
\textbf{MBE [Ours]} & $L=7$ & 100 & 50  & 200 & \textbf{0.971} & \textbf{50.0\%}\\
\bottomrule
\end{tabular*}
\end{ruledtabular}
\end{table*}

\begin{table*}[htbp]
\caption{\label{tab:table2}
\textbf{Comparison of single-graph MBE with circuits of depth $L=13$ and the leading single-shot classical relaxation heuristic~\cite{Burer2001}}, with comparable parameters, for $n=512$ vertex graphs. MBE produces improved solutions (higher cut $\mathcal{C}$), both on average and in the most successful run.
}
\centering
\begin{ruledtabular}
\begin{tabular*}{1\textwidth}{@{\extracolsep{\fill}} ccccc}
\toprule
\textbf{Method}&
\textbf{Mean($\mathcal{C}$)/MaxCut($G$)}&
\textbf{Max($\mathcal{C}$)/MaxCut($G$)}
\\
\midrule
\textbf{Classical Relaxation} & 0.939 & 0.969\\
\textbf{MBE [Ours]} ($L=13$) & \textbf{0.948} & \textbf{0.978}\\
\bottomrule
\end{tabular*}
\end{ruledtabular}
\end{table*}

As minimizing Eq.\ \ref{eq.Ldense} under the constraints of Eq.\ \ref{eq.normalization} cannot yield classical solutions to Eq.\ \ref{eq.MaxCut}, we define a rounding proceedure for the classification and scoring of a cut $\mathcal{C}$ for a graph $G$:

\begin{equation}
\begin{split}
\mathcal{C}_\text{MBE}(\hat{\theta}; G) = & \sum_{j < i}^{n/2} \frac{w_{ij}^{zz}}{2} \left[1 - R(\langle \sigma_i^z \rangle) R(\langle \sigma_j^z \rangle ) \right] + \\ & \sum_{j < i}^{n/2} \frac{w_{ij}^{xx}}{2} \left[1- R(\langle \sigma_i^x \rangle ) R(\langle \sigma_j^x \rangle ) \right] + \\ & \sum_{i,j}^{n/2} \frac{w_{ij}^{zx}}{2} \left[1- R(\langle \sigma_i^z \rangle ) R(\langle \sigma_j^x \rangle ) \right],
\label{eq.Cdense}
\end{split}
\end{equation}

\noindent where the classically implemented function $R$ rounds the measured expectation values to $\pm 1$. We note that this scoring is our true, or \textit{computational} MaxCut estimate, as it is the MaxCut assignement which results from projecting the qubit measurements of our quantum state from the $[-0.76, 0.76]$ codomain of our linear programming relaxation ($\tanh(x)$ activation function) back into the $\pm 1$ codomain of MaxCut nodes.

\section{Results}

In this section, we empirically validate our approach's performance by solving the MaxCut problem on a divese set of nonlocally connected graphs with up to $512$ vertices. We first introduce the experimental settings and implementation details before presenting the results for two scenarios: i) using MBE to solve $n$-vertex MaxCut problems with only $n/2$ qubits, and ii) using MBE to encode two separate MaxCut graph instances in a single circuit. In addition to having an inherently lower quantum hardware overhead and measurement complexity, both implementations of MBE demonstrate superior optimization performance.

Fig.\ \ref{fig.4b} illustrates the average performance (ratio of cut obtained with largest known solution) of both MBE and VQE circuits for graphs of $n=8, 100$ vertices and the MBE circuit alone for $n=512$. The $n=512$ graph with traditional VQE was too memory inefficient for evaluation on a single NVIDIA A100 GPU. The simulations were completed using TensorLy-Quantum, which runs on a PyTorch \cite{pytorch} backend and implements tensor contractions with Opt-Einsum \cite{opteinsum}. The $n=8$ instances are complete (all-to-all, $n(n-1)/2$-edge) graphs for which we calculated the exact ground truth through brute force computation, the $n=100$ graphs are the first three $0.9$ density weighted ($4455$-edge) MaxCut graphs (cataloged as the w09-100 instances) from the extensively studied Biq Mac library \cite{Wiegele2007}, and the $n=512$ graph  is the pm3-8-50 instance of the DIMACS library \cite{DIMACS}. While the pm3-8-50 graph is relatively sparse ($1536$ edges), it is nonlocally connected. Like other recent works \cite{Patti2020, Dborin2021}, we implement simple entanglement-based pre-training prior to the MBE algorithm (details in the Supplementary Information \cite{SuppInfo}). Shallow circuits of depth $L=7$ ($n=8$ and $n=100$ graphs) and $L=13$ ($n=512$ graph) are selected in order to adopt a protocol suitable for near-term quantum devices, however the performance of the larger graphs ($n=100, 512$) increases with moderately deeper circuits.

MBE consistently demonstrates a $5\%$-$7\%$ average performance increase across all $n$, as seen in Fig.\ \ref{fig.4b}. We emphasize that not only is the MBE algorithm more accurate than traditional VQE, it simultaneously solves MaxCut$(G)$ with \textit{half} the required qubits and parameters, as summarized in Table \ref{tab:table1}. As quantum state space scales exponentially in $n$, this factor of two reduction in required qubits remains significant for quantum computing at scale. Even with very shallow circuit-depth ($L$ increasing only sublogarithmically in $n$ compared to the $100$-vertex BiqMac graphs), MBE outperforms the leading single-shot classical algorithm (Table \ref{tab:table2}) for the $512$-vertex DIMACS graph, achieving an average cut of $\sim 95\%$ of the largest known solution \cite{Festa2010}. MBE also outperforms the classical algorithm in terms of the largest cut obtained for any given run, with $\sim 98\%$ accuracy from just thirty total runs compared to $\sim 97\%$ accuracy from one-hundred total runs. These performance increases would be even greater for deeper circuits, however our current contraction algorithm yields a maximum MBE circuit depth of $L=13$ for $512$-vertex graphs on a single GPU. As the simulation of these networks are ultimately memory-bound, with memory requirements growing exponentially with circuit-depth, effective implementations of the algorithm are not classically tractable at-scale. The simulation of deeper circuits could be provided by tensor contraction backends with improved memory management, such as the cuTensor library, while implementations of this scale on quantum hardware is consistent with the projections for moderate-term quantum devices. Although computational benchmarking for optimization problems has been demonstrated for thousands of qubits \cite{huang2019}, to our knowledge, MBE with $n=512$ is the largest simulation of successful quantum optimization algorithms on nonlocally connected graphs yet conducted.

MBE's improved performance on optimization problems is due to the two-axis constraint on each qubit, which only permits convergence to local minima that are \textit{bistable} points for both the $z$ and $x$-axes. This is in contrast with the monostable condition of traditional VQE. Convergence to a local minima with bistability requires the concurrence of a zero gradient for both independently parametrized axes at a single, non-optimal point in parameter space. As $\mathcal{L}_\text{MBE}$ is best extremized by larger $\langle \sigma^{\zeta} \rangle$, the circuit will tend towards satisfying the equality in Eq.\ \ref{eq.normalization}. As this corresponds to entanglement-free qubits, there is a systematic disentanglement of the circuit into product states throughout training (Fig.\ \ref{fig.4b}, right). To understand this process, note that for the general wavefunction

$$|\phi \rangle = \alpha|0_i 0_r \rangle + \beta|0_i 1_r \rangle + \gamma|1_i 0_r \rangle + \delta|1_i 1_r \rangle$$

\noindent describing any two qubits $i$ and $r$, the lefthand side of Eq.\ \ref{eq.normalization} for qubit $i$ can be written as

\begin{equation}
\begin{split}
\langle \sigma_i^z \rangle^2 + & \langle \sigma_i^x \rangle^2 =  \\ & \left[ (\beta+\gamma)^2 + (\alpha-\delta)^2 \right] \left[ (\beta-\gamma)^2 + (\alpha+\delta)^2 \right].
\end{split}
\end{equation}

\noindent In this form, we note that Eq.\ \ref{eq.normalization} is maximized when the concurrence (entanglement \cite{Wootters2001, Gao2008}) is minimized and vice versa, driving the wave function towards product states as training progresses. Once disentanglement nears completion, the equality in Eq.\ \ref{eq.normalization} begins to hold and for any $\theta_t$ and qubit $i$, such that

\begin{equation}
\langle \sigma_i^z \rangle g_t(\sigma_i^z) = - \langle \sigma_i^x \rangle g_t(\sigma_i^x),
\end{equation}

\noindent where $g_t$ are the gradients as given by Eq.\ \ref{eq:gradient}. As $\langle \sigma^{\zeta}_q \rangle = 0 $ is unfavorable for the optimization of $\mathcal{L}_\text{MBE}$, both axes of each qubit $i$ must be bistable with respect to each angle $\theta_t$ in order for update of that parameter to halt.

In this manner, MBE is a sort of quantum analog to alternating minimization in classical algorithms \cite{Jain2017}, but which uses both quantum superposition and classical nonlinearity to minimize two cost functions simultaneously, rather than one sequentially. Alternating minimization has also proven useful in QAOA protocols \cite{Hadfield2019, Zhu2020, Cook2020, Wang2020Rieffel}, as has other perturbations, such as filtered measurements \cite{Amaro2021}. Because $\mathcal{L}_\text{MBE}$ is calculated from single-qubit measurements, it is a form of measurement-based quantum computation \cite{Raussendorf2001, Raussendorf2003, Ferguson2021}.  Moreover, as the number of possible single-qubit measurements scales linearly with circuit width, $\mathcal{L}_\text{MBE}$ represents up to a quadratic reduction in the number of observables required to solve complete graphs from $\sim n^2$ (specifically $n(n-1)/2$ two-operator Pauli strings) to $\sim 2n$ (two single-qubit measurements per qubit), lowering the measurement complexity and runtime of the algorithm on real quantum hardware \cite{Verteletskyi2020, Shehab2019}.

\begin{figure}[h!]
\subfloat[(Left) The probability $P(\mathcal{C} > T)$ that cut $\mathcal{C}$ of an $n=100$ graph is optimal using: MBE with $L=13$ (light green),  MBE with $L=7$ (dark green), and VQE with $L=7$ (black). Increasing depth from $L=7$ to $L=13$, while still shallow for $n=100$, markedly improves performance. (Right) $P(\mathcal{C} > T)$ of $n=100$ graphs using: two-graph MBE with $L=7$ (light green), two-graph MBE with $L=1$ (dark green), and VQE with $L=7$ (black). While the $L=1$ case is entanglement-free, it benefits from MBE's two-axis constraints. \label{fig.3a}]{%
  \includegraphics[width=\columnwidth]{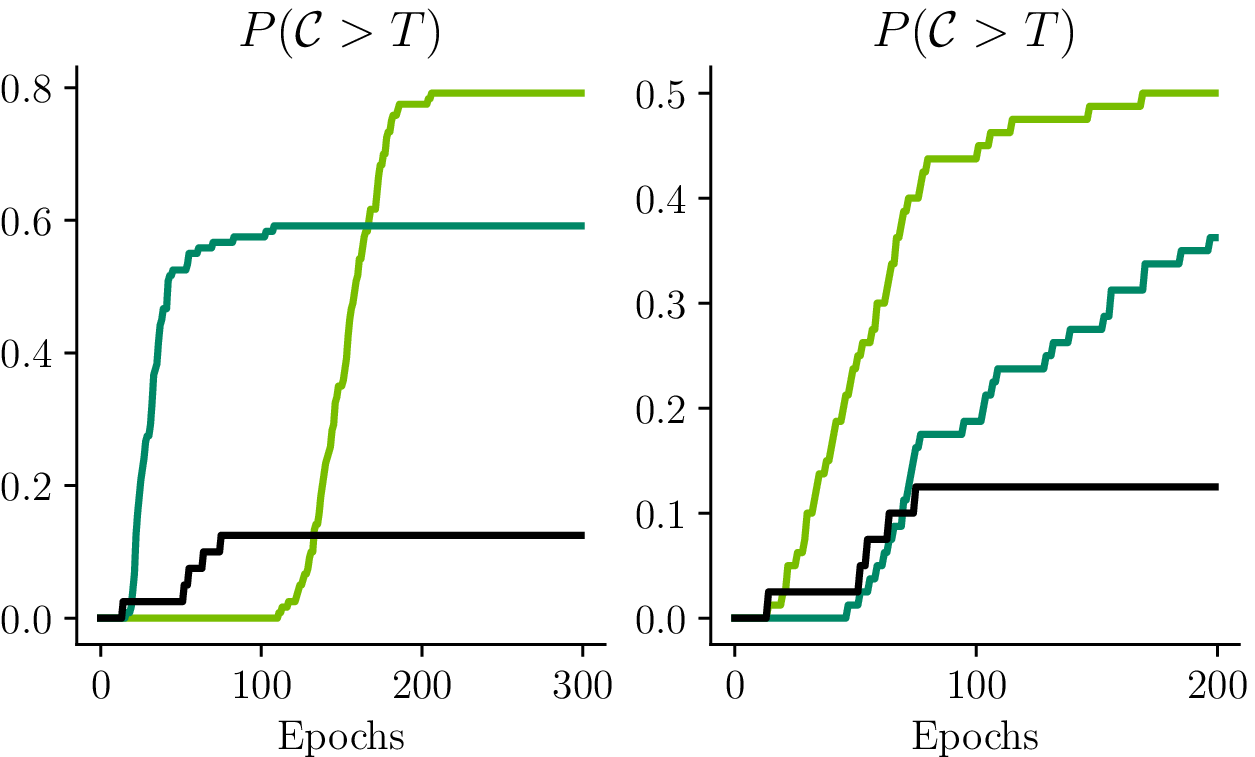}%
}\hfill
\subfloat[(Left) The probability of achieving an optimal cut ($\mathcal{C} > T$) of an $n=100$ graph with $r=5$ repeats using:  two-graph MBE with $L=7$ (light green), two-graph MBE with $L=1$ (dark green), and VQE with $L=7$ (black). For the shallow $L=7$ MBE circuit, five repetitions produces nearly deterministic results with less than $200$ epochs. (Right) Number of $n=20$ graphs with identified optimal cuts from set of ten instances and $r=10$ repeats using: two-graph MBE (green), and VQE (black). MBE not only successfully optimizes all (vs $90\%$) of $G$, it solves twice as many graphs in the same number of epochs. \label{fig.3b}]{%
  \includegraphics[width=1.05\columnwidth]{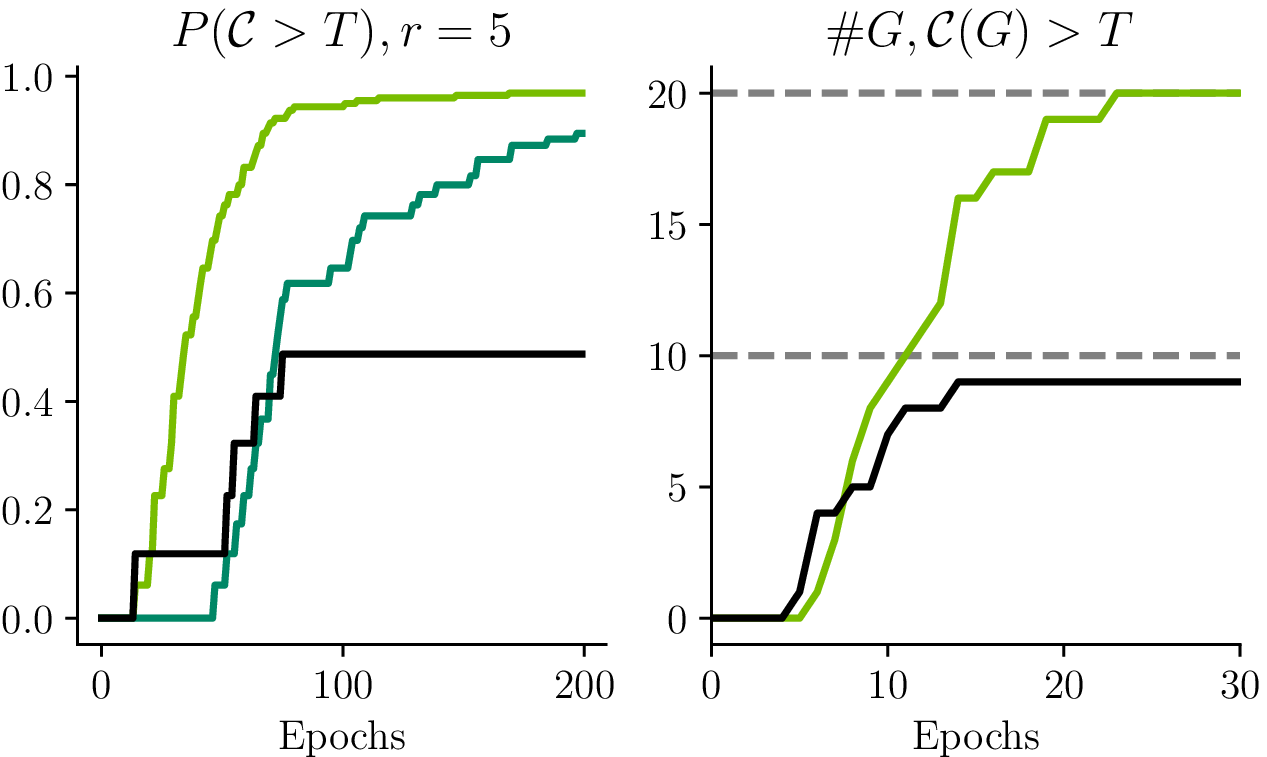}%
}
\caption{}
\label{}
\end{figure}

MBE can also encode two \textit{distinct} $n$-vertex graphs into a single register of $n$-qubits and solve their two MaxCuts in parallel. This is equivalent to the simplified case of $w_{ij}^{zx}=0$ $\forall i,j \leq n$ in Eqs.\ \ref{eq.Ldense} and \ref{eq.Cdense} using $n$ qubits, yielding

\begin{equation}
\begin{split}
\mathcal{L}_\text{MBE} = & \sum_{j < i}^{n} w_{ij}^{zz} \tanh(\langle \sigma_i^z \rangle ) \tanh(\langle \sigma_j^z \rangle ) \\ & + \sum_{j < i}^{n} w_{ij}^{xx} \tanh(\langle \sigma_i^x \rangle ) \tanh(\langle \sigma_j^x \rangle ),
\label{eq.Lparallel}
\end{split}
\end{equation}

\noindent and

\begin{equation}
\begin{split}
\mathcal{C}_\text{MBE}(\hat{\theta}; G) = & \sum_{j < i}^{n} \frac{w_{ij}^{zz}}{2} \left[1 - R(\langle \sigma_i^z \rangle) R(\langle \sigma_j^z \rangle ) \right] \\ & +\sum_{j < i}^{n} \frac{w_{ij}^{xx}}{2} \left[1- R(\langle \sigma_i^x \rangle ) R(\langle \sigma_j^x \rangle ) \right].
\label{eq.Cparallel}
\end{split}
\end{equation}

\noindent The average performance of MBE for solving two $n$-vertex graphs in parallel vs that of traditional VQE with a single graph is displayed in Fig.\ \ref{fig.2b} for graphs of $n=8$, $20$, and $100$ vertices with $L=7$. For $n=8$ and $n=20$, we generate exact solutions to complete (all-to-all) graphs through brute force computation, whereas the $n=100$ graphs are again the first three $0.9$ density weighted MaxCut graphs from the  Biq Mac library \cite{Wiegele2007}. While for this fixed $L$, both VQE and two-graph MBE suffer decreasing performance with increasing $n$, two-graph MBE consistently demonstrates a $5\%$-$7\%$ average performance increase across $n$. We again note that the performance for large-$n$ graphs increases with greater $L$. Finally, we emphasize that not only is the MBE algorithm more accurate than traditional VQE, it simultaneously solves MaxCut$(G)$ for \textit{two} graphs $G$, rather than only one as with traditional VQE.

Although much emphasis is placed on the development of quantum algorithms that deterministically obtain optimal cuts, studies have indicated that this requires up to an exponential number of parameters with traditional VQE \cite{lee2021favorable}. This is an unfeasible quantity, reaching $\sim 2^{99}$ ($\sim 2^{511}$) parameters for the $n=100$ ($n=512$) graphs considered here. Conversely, the cumulative effects of probabilistic sampling (that is, running the randomly initialized circuit multiple times) lead to high-confidence convergence with markedly few repetitions $r$. In what follows, we reason that a probabilistic sampling of various shallow MBE circuit initializations is a more efficient alternative. As larger values of $\mathcal{C}$ are a direct certificate of superior optimization, there should be no preference for less efficient single-shot techniques. Furthermore, shallow implementations are particularly important for near-term quantum devices, which are prohibitively susceptible to noise at even moderate circuit-depth.

Fig.\ \ref{fig.3a} displays the probability that an optimal cut, which we define as $\mathcal{C} > T = 0.97 \times \textrm{MaxCut}(G)$, will be found for $n=100$ graphs with both MBE and VQE. For depth $L=7$, MBE produces an optimal cut with upwards of $50\%$ probability for both the single-graph ($n$ vertices in $n/2$ qubits, Fig.\ \ref{fig.3a} left)  and double-graph (two $n$ vertex graphs in $n$ qubits, Fig.\ \ref{fig.3a} right) protocols. In contrast, traditional VQE with $L=7$ produces optimal cuts with just $12.5\%$ probability. Furthermore, the likelihood of obtaining an optimal cut with MBE increases considerably with moderate circuit depth, rising to approximately $80\%$ for $L=13$ (left). We note that $L=1$ circuits (right) obtain optimal cuts with probability $0.36$, tripling the convergence rate of standard VQE with $1/7$th the resources. As circuits with $L=1$ are comprised of only local rotations without control gates, the totality of the performance is due to mutual constraints on multi-basis \textit{superpositions}, and not due to quantum entanglement. Like other entanglement-free formulations \cite{Wang2019, Goto2019, Crosson2016}, this renders the circuit efficient for classical simulation and indicates that algorithms for simulated superposition with multi-basis constraints may hold promise as ``quantum inspired" classical algorithms. However, we note that quantum implementations are still of interest, because other entanglement-free relaxations are known to suffer decreased performance with increasing circuit width $n$ \cite{lee2021favorable}. Furthermore, MBE with even modest entanglement and circuit-depth markedly increases the probability of optimal convergence.

Fig.\ \ref{fig.3b} (left) shows the probability of obtaining at least one optimal cut for $n=100$ graphs with $L=7$ and $r=5$, which nears $97\%$ in fewer than $100$ training steps for two-graph MBE circuits. For $r=10$, convergence is greater than $99.9\%$ and the $4nr=4000$ parameters utilized for ten repetitions still pale in comparison to the exponentially many required by deep-circuit techniques. As traditional VQE with $L=7$ and $n=100$ produces optimal cuts only $12.5\%$ of the time, MBE is four times more effective than VQE for probabilistic optimization.

MBE also offers superior performance over traditional VQE in terms of the diversity of tenable graphs (Fig.\ \ref{fig.3b}, right). For $r=10$, not only does two-graph MBE find optimal solutions for \textit{all} of the complete $n=20$ graphs tested (compared to $90\%$ for VQE), its parallel implementation doubles the number of MaxCut instances optimized.

\textbf{Simulation Considerations -}
Numerically, $\mathcal{L}_\text{MBE}$ is more compact for large or dense graphs, where the MPO $H$ quickly becomes cumbersome. However, for the single-qubit measurements required for $\mathcal{L}_\text{MBE}$, contraction with a simple, single-qubit operator needs to occur $n$ times. In order to efficiently compute $n$ single-qubit measurements on large, exact tensor networks without either reconstructing an exponentially large ($2^{n/2}$) space or contracting over the full network $\sim n$ times, we use an efficient partial trace-based contraction scheme in which we construct $k$ distinct reduced density matrix operators

\begin{equation}
\rho_k = \sum_{\{\beta, \gamma, \delta \notin K\}} \Psi^{\{\beta\}} U^{\{\beta, \gamma\}} U^{\{\gamma, \delta\}} \Psi^{\{\delta\}},
\end{equation}

\noindent where $K$ is the $k$th set of kept indices. $K$ should be sufficiently small so that the $2^{|K|}$ elements of $\rho_k$ remain numerically tractable. For each $\rho_k$, $|K|$ smaller partial traces are done to isolate single-qubit density matrices $\rho_q$, with which we take the single-qubit expectation values of Eq.\ \ref{eq.Lparallel}

\begin{equation}
\langle \sigma^{\zeta}_q \rangle = \textrm{Tr} \left[ \sigma^{\zeta}_q \rho_q \right],
\end{equation}

\noindent where $\zeta = z, x$.

\section{Discussion}

In this manuscript, we introduced Multi-Basis Encoding (MBE), a novel technique for quantum optimization algorithms. MBE's performance on a diverse set of graphs exceeds that of traditional VQAs. MBE also provides meaningful efficiency improvements over similar VQAs, potentially closing the gap between near-term implementations and quantum advantage by reducing the overhead of quantum algorithms. These efficiency improvements include up to a quadratic reduction in circuit measurements, as well as a factor of two decrease in required qubits, which can readily be extended to a factor of three with the inclusion of the $y$-basis. While simulated using classically tractable ansatze, the performance of our algorithm benefits from increased circuit-depth. As the classical simulation complexity increases exponentially in circuit-depth, this indicates that MBEs may enjoy meaningful quantum advantages at-scale. Furthermore, when we extend our definition of accuracy to encompass probabilistic sampling of various circuit initializations, we find that remarkably few quantum resources are requisite for classical optimization problems.

MBE can be expanded to a broad framework of multi-axis qubit encodings, which would include any nonlinear quantum loss function that permits the optimization of multiple, mutually regularizing observables on a single qubit. These findings are likely to spur additional research in efficient qubit encodings and the application of our techniques to related algorithms. These include algorithms with high circuit-depth or high circuit-connectivity, which are intractable on classical hardware and thus represent clear opportunities for quantum advantage. Since deeper circuits are attainable with more efficient tensor contraction methods or distributed computing efforts, this work encourages further development of large-scale quantum simulation with tensor methods. Most critically, as these simulations are ultimately memory-bound, the implementation of MBE at-scale constitutes a strong and novel candidate for quantum advantage.

We also leverage the powerful tensor techniques packaged in TensorLy-Quantum to complete large-scale simulations of effective optimization algorithms on a single, consumer-grade GPU. To our knowledge, we have produced the largest to-date simulation of a quantum algorithm for a nonlocally connected optimization problem that rivals classical performance. Such a successful and large-scale implementation demonstrates that simple and low-rank tensor representations are sufficient to model various techniques in quantum machine learning, and to do so without truncation or approximation. Finally, through the use of large-scale nonlocally connected graphs, we demonstrate that the global qubit connectivity and high entanglement capacity lacked by both the MPS formalism and linearly connected near-term quantum devices do not preclude quantum optimization routines.\\\\

\section{Acknowledgements}

This work was done during T.L.P.'s internship at NVIDIA. At CalTech, A.A. is supported in part by the Bren endowed chair, and Microsoft, Google, Adobe faculty fellowships. S.F.Y. thanks the AFOSR and the NSF for funding. The authors would like to thank Brucek Khailany, Johnnie Gray, Garnet Chan, Andreas Hehn, and Adam Jedrych for conversations.

\bibliography{MyCollectionTrunc}

\end{document}